# Effect of broken axial symmetry on the electric dipole strength and the collective enhancement of level densities in heavy nuclei

E. Grosse,
Institute of Nuclear and Particle Physics, Technische Universität Dresden,
01062 Dresden, Germany

A.R. Junghans,
Institute of Radiation Physics,
Helmholtz-Zentrum Dresden-Rossendorf, 01314 Dresden, Germany

J.N. Wilson,
Institute of Nuclear Physics,
Université Paris-Sud and CNRS/IN2P3, 91406 ORSAY, France

**Abstract**
The basic parameters for calculations of radiative neutron capture , photon strength functions and nuclear level densities near the neutron separation energy are determined based on experimental data without an ad-hoc assumption about axial symmetry - at variance to previous analysis. Surprisingly few global fit parameters are needed in addition to information on nuclear deformation, taken from Hartree Fock Bogolyubov (HFB) calculations with the Gogny force, and the generator coordinator method (GCM) assures properly defined angular momentum. For a large number of nuclei the GDR shapes and the photon strength are described by the sum of three Lorentzians (TLO), extrapolated to low energies and normalized in accordance to the dipole sum rule. Level densities are influenced strongly by the significant collective enhancement based on the breaking of shape symmetry. The replacement of axial symmetry by the less stringent requirement of invariance against rotation by 180 degree leads to a novel prediction for radiative neutron capture. It compares well to recent compilations of average radiative widths and Maxwellian average cross sections for neutron capture by even target nuclei. An extension to higher spin promises a reliable prediction for various compound nuclear reactions also outside the valley of stability. Such predictions are of high importance for future nuclear energy systems and waste transmutation as well as for the understanding of the cosmic synthesis of heavy elements.

Corresponding author : E.Grosse@tu-dresden.de

# I. Introduction

The ongoing discussion, e.g. in this volume of Physica Scripta, about shapes and shape changes in heavy nuclei, often concerning triaxial shapes far off stability, induces the question, how well the widely used *ad hoc* assumption about axial symmetry [1] in less exotic nuclei is based on sufficiently sensitive experimental results. Here experimental data for masses and level energies seem to be not very conclusive; thus it is appealing to regard other observables with respect to axial symmetry breaking. Triaxiality has long known to be important for the fission process [2, 3] as well as for odd nuclei [4]. At variance, energy spectra observed in even nuclei were for a long time interpreted assuming at least axial symmetry of the nuclear shape [5, 6, 1], There are some more recent instances from gamma ray spectroscopy in heavy nuclei where triaxiality has been claimed to be directly or indirectly observed with either triaxial wobbling phonon bands or the possible presence of chiral-symmetric bands [7]. But for the analysis of the experimental results and for the unambiguous identification of triaxial shapes from the energies of excited nuclear states complex theoretical considerations are needed. A similar statement can be made for multiple Coulomb excitation studies, influenced by axial shape symmetry and its breaking [8]. This had led to stating triaxiality already for some nuclei in the valley of stability [9, 10, 11, 12].

In this paper we present an analysis of more directly observable manifestations of a deviation from axial symmetry. We regard not only energies of single transitions in selected nuclei, but we use very many levels distributed near the neutron separation energy and in the IVGDR range as signals. The splitting of the Isovector Giant Dipole Resonance (IVGDR) is used in text books as an indicator of axial deformation, but apparently an adjustment of several parameters for each individual isotope is needed [13]. In the work presented here this observable is reviewed under the assumption of broken axial symmetry; a description with only global parameters has been shown [14, 15, 16, 17, 18, 19, 20, 21] to be valid for many heavy nuclei in the range of 70<A<200; here we present novel results for the Nd-isotopes. An even stronger reduction of the number of parameters is found for the other topic of this work, the accordance of neutron resonance spacings and other level density data to a Fermi gas prediction, achieved by us without any parameter fitting. Observations obtained with more than 140 spin-0 target nuclei will be interpreted by broken axial shape symmetry. A common feature in the interpretation of these two experimental phenomena is the replacement of the assumption of axiality by the less stringent requirement of invariance against rotation by an angle π, $\mathcal{R}_\pi$.

Our findings have a strong impact on the radiative capture process, for which also the low energy tail of the dipole strength is of major importance, as noted recently for nuclei with mass number A>70 [14]. This process plays an important role in considerations for advanced nuclear systems [22, 23, 24] and devices aiming for the transmutation of radioactive nuclear waste. It also is of interest for the cosmic nucleosynthesis with fluxes of neutrons that high, that their capture reaches heavy nuclides beyond Fe [25, 26]. The experimental studies forming the basis for respective predictions can mainly be performed on nuclei in or close to the valley of beta-stability. Thus a small number of global parameters, as we show to suffice, are of great advantage.

## II. Quadrupole observables and axiality

As basis for the discussion of the IVGDR we will first describe connections between nuclear shapes and electric quadrupole moments and transition rates with and without the assumption of axial symmetry. To demonstrate the importance of electromagnetic sum rules for the nuclear dipole strength, general features of the IVGDR will be presented. A departure of nuclear shapes from spherical symmetry was first indicated by a splitting of atomic transitions due to the form of the nuclear electromagnetic field [27]. Hyperfine structure measurements, improved in accuracy using laser techniques, as well as muonic X-ray studies, determined the 'spectroscopic' electric quadrupole ($\lambda$=2) moment $Q_s$ of the ground state [28] in nearly 800 odd nuclei. In addition, the reorientation effect in Coulomb excitation made $Q_s$-values also available for excited $2^+$ and $4^+$-states $Q_s(I^\pi)$ in even nuclei. Mostly, the sign of $Q_s(2^+)$ was determined to be negative [28] and the observed $Q_s$ were often not in agreement to predictions for single particle or hole configurations [1]. In a semiclassical picture of collective rotation [5] the intrinsic structure and shape for the ground state $0^+$ and the lowest $2^+$-state $r$ are assumed to be the same. For $Q_s<0$ this picture suggests an apparently oblate shape to result from the rotation of a more prolate body with an 'intrinsic' quadrupole moment $Q_i = -\frac{7}{2} Q_s$. Quantum mechanically the rotation about a symmetry axis is forbidden, and a projection on proper angular momentum in the laboratory frame is necessary. With a homogeneous distribution of the charge within the nuclear volume, the intrinsic electric quadrupole moment $Q_i$ of even nuclei is related to the difference in half-axis length $\Delta R$ between the symmetry axis of the shape $R_3$ and the two short ones (in case of axial symmetry $R_1=R_2$) by setting [5]:

$$Q_i \equiv \sqrt{\frac{9}{5\pi}} ZR^2 \beta (1 + b \cdot \beta) \, ;$$

$$\beta \cong \frac{4}{3}\sqrt{\frac{\pi}{5}} \frac{\Delta R}{R} \cong 1.057 \frac{\Delta R}{R} \qquad (1)$$

For the axial case a single deformation parameter $\beta$ was introduced and the relation (1) between deformation $\beta$, $Q_0$ and $\Delta R$ is widely applied when electromagnetic data are related to calculated nuclear (mass) deformations, usually characterized by $\beta$. For years $b \approx 0.16$ was assumed [5, 29], but later an often used compilation of electric quadrupole transition widths [30] proposed $b = 0$ as approximation. Besides this ambiguity in b several definitions proposed as deformation parameters in the literature [31, 8, 30, 32] may differ from the 'standard' definition as given here. Another observable for a deformation of nuclei is the splitting of the IVGDR and there the axis lengths are the quantity of importance.

The enhancement seen in experimental data on electric quadrupole (E2) transitions from the ground state [30] indicates a strong excess above predictions for a transition to a configuration formed by exciting only few particles. In connection to the observed quadrupole moments mentioned above this was linked [1] to the breaking of spherical symmetry in quasi all heavy nuclei away from magic shells. The model of a rotating axially symmetric liquid drop with a quadrupole moment, representing an even nucleus, predicts one rotation related $2^+$-state with a 'collective' i.e. enhanced E2-transition width. Then $Q_i$ (in fm$^2$) is related to the reduced matrix elements (in e·fm$^2$) for an electric quadrupole transition from this 'rotational' state $r$ to the ground state [5, 1] by $E_\gamma$ (in MeV):

$$\frac{5}{16\pi} Q_i^2 = |\langle r\|\mathbf{E2}\|0\rangle|^2 = B(E2, 0 \to r) \quad (2).$$

The E2 ground state decay width $\Gamma_{r0}$ ($E_\gamma$) (in MeV) is obtained from the general relation:

$$\Gamma_{r0}(E_\gamma; E2) = \frac{4\pi}{75} \frac{\alpha_e E_\gamma^5}{g(\hbar c)^4} |\langle r\|\mathbf{E2}\|0\rangle|^2 ;$$

$$g = \frac{2J_r+1}{2J_0+1} \qquad (3)$$

where $\alpha_e$, $\hbar$ and c are the fine structure constant, Planck's constant and the velocity of light; $J_0$ and $J_r$ are the spins of the ground state and the excited level. The reduced transition probability B(E2) used in Eq. (2) describes the quadrupole transition between the ground and the lowest $2^+$-state. $Q_s$ and B(E2) are observables, whereas the deformation $\beta$ is a model parameter, and in the relation (1) to data one assumes a uniform axial charge distribution. One serious shortcoming of the axial rigid rotor model is the fact, that it only predicts one 'collective' $2^+$-state. Experimentally at least two $2^+$-levels with enhanced transitions to the ground state are observed in nearly all even nuclei. Then a sum of all ground state transitions appears [8] in Eq. (2). One possible explanation [29, 1] is the coupling of the nuclear rotation to a collective quadrupolar vibration around an axially deformed basis state.

As an alternative origin of a second low energy $2^+$-state a static triaxiality with the possibility of more than one rotation axis has been regarded [33]. For a confirmation we now refer to self-consistent microscopic calculations as a representation of the nucleus as an ensemble of Z+N=A nucleons in a compact volume V. We only regard calculations which are not based 'ad hoc' on the assumption of axiality and mention that already long ago it was pointed out [34], how a projection from the intrinsic system into the observer's frame quasi automatically leads to triaxiality as result of a Hartree-Fock-Bogoliubov (HFB) calculation. As required by quantum mechanics, such a projection has to be made after the HF-variation, and this has the consequence that the expectation value for the triaxiality, i.e. the $\gamma$-mode, is different from zero: $\langle\gamma\rangle \neq 0$; semi-classically this is equivalent to a $\gamma$-oscillation centred at a finite $\gamma$. A first calculation [34] was performed for two heavy nuclei only, but a more recent one [32] is available for practically all heavy even nuclei between the neutron and proton drip lines. These constrained HFB calculations "*are free of parameters beyond those contained in the Gogny D1S interaction*" [32]. Assuming only $\mathcal{R}_\pi$-invariance they find non-zero triaxiality for many nuclei, and in some cases the predicted standard deviation does not include $\gamma = 0$. The use of constrained wave functions and a generator coordinate method allowed to project on good angular momentum as proposed before [34]. In view of an agreement to our earlier experimental findings [18, 16, 35, 19] we rely on these calculations and use the predicted $\gamma$–values. Following a suggestion made in an additional HFB-study [36] we reduce the $\beta$–deformation parameters: for nuclei which are only $\delta$ nucleons away from a shell (with $\delta \leq 10$) a factor $\varepsilon = 0.4 \pm \delta$ is applied to obtain $\beta_{\text{eff}} = \varepsilon \cdot \beta$. The paper by Delaroche et al. [32] has an attachment which lists values for $\beta$ and $\gamma$ for more than 1700 nuclei as well as radius parameters, all derived from CHFB+GCM. From $\beta_{\text{eff}}$, $\gamma$ and the proton radius $R_p$ three half-axes $R_1$, $R_2$, $R_3$ for each nucleus were extracted; here we used their Eq. (3) for the three oscillator parameters, which are inversely proportional to the half-axes. This invokes the concept of an equivalent ellipsoid, which has the same charge Ze, volume V and quadrupole moment $Q_i$ as the nucleus. Using Eqs. (19-21) in the work of Kumar [8] (correcting for a missing factor of 2) the intrinsic 'collective' $Q_i$ can be obtained by:

$$Q_i = \frac{2Z}{5} \cdot \sqrt{(2R_3^2 - R_1^2 - R_2^2)^2 + 3(R_1^2 - R_2^2)^2}$$

$$R_p^3 = R_1 \cdot R_2 \cdot R_3 = \frac{3\pi}{4} V \qquad (4)$$

Here we assume only $\mathcal{R}_\pi$–invariance as well as identical distributions of protons and neutrons. In Fig. 1 the correlation between $\gamma$ and $Q_i$ is depicted for nuclei in the minimum of the valley of stability and $Z\pm 1$.

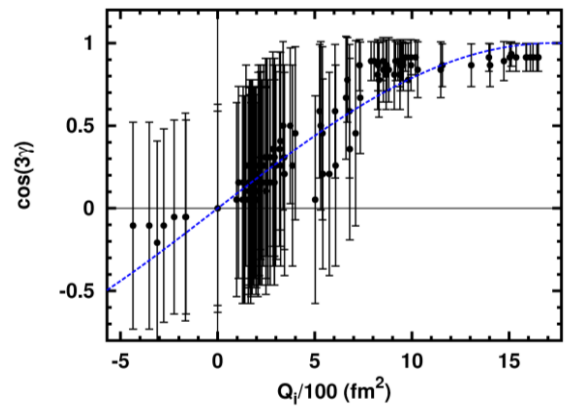

***Fig. 1:*** *Correlation between cos(3γ) and $Q_i$ in ≈ 130 even nuclei with 60<A<250; the respective data are taken from a CHFB+GCM calculation [32] for nuclei in the valley of stability. The bar lengths represent the standard deviations in γ as given by these calculations and tabulated as supplemental material.*

In the rotation invariant ansatz [8] the deviation from axial symmetry is described by the parameter cos(3γ), which also defines the sign of $Q_i$ and which we use in Fig. 1. As in quantum-mechanical systems like nuclei only expectation values are accessible to measurements, cos(3γ) and $Q_i$ in the Figure are to be understood as such. In Fig. 1 a clear trend to triaxiality with decreasing $Q_i$ is obvious (cos(3γ)→0), whereas most well deformed nuclei show a clearly smaller deviation from axiality. The small number of nuclei which are oblate already at low $E_x$ (cos(3γ) < 0) does not allow similar conclusions, and for very small $Q_i$ a triaxiality would be very difficult to distinguish from sphericity. The trend as indicated as blue dashed curve in Fig. 1 suggests an approximation of nuclear shapes by only one parameter $Q_i$, with axiality depending on it. The clustering at $Q_i$ < 200 fm$^2$ and cos(3γ) ≲ 0.2 seen in Fig. 1 is significant and will play an important role for the discussion of IVGDR shapes in the numerous nuclei with intermediate $Q_i$, often called 'transitional'.

### III.  Photon strength and sum rules

From very general conditions like '*causality and analyticity*' together with dispersion relations the Thomson scattering cross section was generalized by quantum electro-dynamics [37] to shorter wavelength photons interacting with nuclei of mass number A=Z+N. This lead to the Gell-Mann-Goldberger-Thirring (GGT) sum rule, predicting the cross section for the absorption of photons by nuclei, integrated up to the threshold for sub-nuclear processes:

$$\int_o^{E_u} \sigma(E_\gamma) dE \lesssim \frac{2\pi^2(\alpha\hbar^2)}{m_N}[ZN/A + A/10]$$
$$\approx 5.97\,[ZN/A+A/10]\ \text{MeV fm}^2 \quad (5)$$
$$E_u = m_\pi c^2$$

Here $m_N$ and $m_\pi$ stand for the mass of nucleon and pion, respectively. The first term in the sum is the "classical sum rule" of Thomas, Reiche and Kuhn (TRK, [38]) and the overshoot over it predicted and discussed [13] is contained in the second. This term has been shown to be accurate within 30% as approximated by assuming [39] "*that a photon of extremely large energy interacts with the nucleus as a system of free nucleons*", and only above an upper energy $E_u$ hadronic degrees of freedom become important. Eq. (5) comprises all multipole modes of photon absorption and includes the absorption by nucleon pairs and especially p-n-pairs, which are strongly dissociated by photons with 20<$E_\gamma$<200 MeV. The respective "quasi-deuteron effect" has been derived from the expression valid for the free deuteron by correcting for Pauli blocking [40].

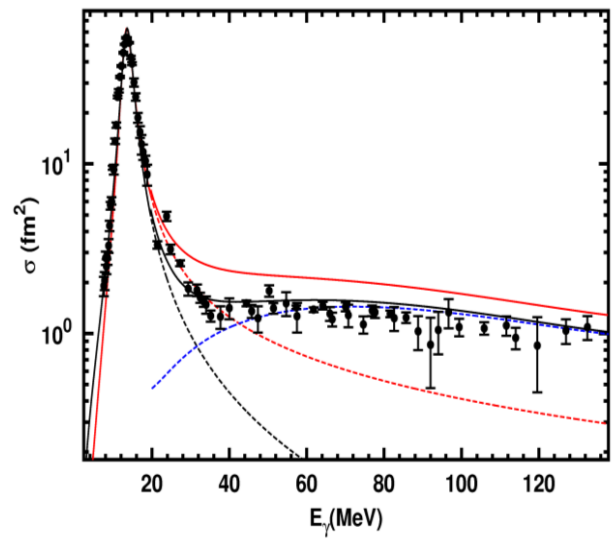

***Fig.2***: *Cross section of photo-neutron production on $^{208}$Pb[41] in comparison to a Lorentzian for the isovector IVGDR (black and red dashed lines, see text) and the quasi-deuteron effect (blue dashed line). The sum of both contributions is given as drawn lines.*

Photo-neutron data are available [41] for $^{208}$Pb up to energies above $m_\pi c^2$; they are shown in Fig.2 and compared on an absolute scale to a Lorentzian given by the first term in Eq. (5). The integral of this term agrees to the TRK sum, and also the expression for the absorption cross section corresponding to the quasi-deuteron mode [40] for $E_\gamma$ > 20 MeV is given on absolute scale. The sum of both is depicted as well and the case of a constant width $\Gamma_r$ is shown in black. The change which evolves from making it proportional [13] to the photon energy $\Gamma_r \propto E_\gamma/E_r$ is demonstrated in red, but obviously the data above 25 MeV are clearly below this curve, and the disagreement of such a change to the data shows that the proposed width change with photon energy does not hold above the IVGDR. Another approach known as KMF model [42] was quoted to evolve from Landau theory of Fermi liquids and even proposes to make the width proportional to the square of the photon energy. A theoretical work [43] finds Fermi liquid theory not applicable to E1 modes in nuclei, such that the disagreement expected to become even greater in a comparison for $^{208}$Pb in this energy range; the situation is similar for the various other nuclei studied [40].

Fig. 2 depicts a width $\Gamma_r$=3 MeV and pole energy $E_r$=13.6 MeV, which were both predicted for the IVGDR [44] by an older HFB-calculation. The slight enhancement over the sum of the two curves, obvious in the figure close to 25 MeV, may be assigned to the Isovector Giant Quadrupole Resonance. The IVGDR is well described in the region below 20 MeV by the first term in Eq. (5), *i.e.* the classical electric dipole sum rule (TRK). Apparently the absorption above 40 MeV mainly corresponds to the quasi-deuteron mode and its integral is close to the second term in Eq. (5). Our description using three Lorentzians (TLO) will be discussed now for nuclei away from the doubly magic $^{208}$Pb.

## IV. Isovector giant dipole resonances

The photo-disintegration of nuclei as one of the first studied nuclear reactions has soon after been recognized as a manifestation of a collective excitation mode [45, 46]. The first theoretical descriptions for the oscillation of protons against neutrons were well describing medium mass nuclei [45] and the very heavy ones [46]. By using the concept of the droplet model these two approaches were unified and IVGDR centroid energies $E_0(Z,A)$ were reasonably well predicted [47] in the range 60<A<240. The predicted IVGDR pole energies were used [ju08] to derive a procedure based on three Lorentzians yielding a global parameterization of the electric dipole strength. Here, we follow that work and a symmetry energy J=32.7 MeV and a surface stiffness Q=29.2 MeV from the finite range droplet model [48] are used, but the nuclear radius is now taken as $R_p$, as in Eq. (4). Only one additional parameter, an effective nucleon mass $m_{eff}$ = 800 MeV, had to be adjusted to give an overall fit to the IVGDR data for 70<A<200. These parameters are combined to predict $E_0$, as was shown previously [14]; the difference in $m_{eff}$ is related to the new choice of R=$R_p$ as taken from the calculations [32].

The splitting of the IVGDR in the deformed lanthanide and actinide nuclei is obvious in the experimental data [48, 49]. Since long, the coupling of dipole and quadrupole degrees of freedom in heavy nuclei has been discussed [50, 51, 29] and detailed calculations [52, 53] within various models have obtained reasonable fits to experimental data for selected nuclei. The parameterization to be presented here is much less ambiguous concerning the mode coupling, but it incorporates nuclear triaxiality explicitly using a description by a sum of k=3 Lorentzians. In heavy nuclei in general, the apparent width of the IVGDR is determined by several components:

(a) Spreading into underlying configurations,

(b) Nuclear shape induced splitting,

(c) Fragmentation and

(d) Particle escape.

From calculations for heavy nuclei using the Rossendorf continuum shell model [54, 55] the escape widths (d) in the IVGDR region were shown to be clearly smaller than the widening caused by damping or spreading as predicted by Eq. (6). For the concept of fragmentation (c) of the configurations belonging to e.g. the IVGDR a calculation of these configurations is needed. The detailed shell model calculation [56, 20] for the nucleus $^{208}$Pb, which is based on a large number of configurations and the experimental energy resolution suggest a smooth description of the data. In addition, a detailed calculation [44] indicates the quality of a parameterization by Lorentzians for the IVGDRs, and this is supported by high statistics data [57] which do not justify a dependence of the width on photon energy. When a photon-energy independent Γ is used, an agreement to data above 25 MeV is reached as well; this was demonstrated in Fig. 2.

In various papers [14, 17, 16, 18, 35] it was shown by the Dresden group that a Lorentzian description is possible also for nuclei away from closed shells with A>60, if proper account is made for the ground state deformations. Using that the resonance width Γ depends on $E_{IVGDR}$ only and thus smoothly on A and Z, it was demonstrated that accord to the classical dipole sum rule is reached to a surprising degree. Here, hydrodynamical considerations [58] predict the dependence of the damping width $Γ_k$ of an IVGDR on its pole energy $E_k$ in good agreement to experimental findings [59]. With one parameter adjusted to be equal for all heavy nuclei with A>70 one gets - if both are expressed in MeV:

$$Γ_k \approx c_w E_k^{1.6} \qquad (6).$$

Of course, the proportionality constant has an uncertainty related to the selection of nuclei, which are included in the fit. With the axis ratios from the CHFB calculations we get $c_w$ =0.045(3) from a fit to nearly all nuclides for which respective data exist [60, 61]; $c_w$ is no longer free, if the width prediction for $^{208}$Pb [44] is transferred to other nuclei. As the slope of a Lorentzian sufficiently far away form $E_0$ is directly proportional to $Γ_k$ its uncertainty directly enters in the radiative width and the large unsystematic scatter seen in the local fits [60, 13] yield strong arguments against their use. When a parameterization of the electric dipole strength in non-spherical nuclei is aimed for, the contribution (b) has to be treated sufficiently well. Lorentzian fits [13] to data performed for each nucleus independently cause a wide fluctuation of the apparent width with Z and A [60]; a non-systematic variation of the damping is difficult to conceive within the spreading concept. A similarly erratic dependence of the integrated IVGDR strength on Z and A was also reported [13] to result from this approach of fitting the photo-absorption data locally. In some cases the integrated cross section overshoots the smooth trend given by Eq. (5, first term, classical sum rule) by up to 100 %. Apparently the two problems named are closely related, as the resonance integral is proportional to the product of height and width. As proposed previously [14, 15, 16, 17], a solution for this problem is found through the incorporation of nuclear triaxiality and this point will now be examined in further detail. In our ansatz the resonance energy $E_0$ is modulated by using the ratios of the $ω_k$, respectively their inverse, the axis lengths $R_k$ from the CHFB+GCM calculation [32] discussed already; this information yields the three energies for the splitting of the IVGDR into three components of equal strength, centred at $E_0$ [14]:

$$E_k = \frac{R_0}{R_k} \cdot E_0 \qquad (7)$$

As will be seen in Fig. 3 the energy splitting between the three IVGDR components is comparable to their widths for many nuclei and thus a triple split should be introduced explicitly; this is especially indicated for nuclides with $Q_i \lesssim$ 200 fm$^2$, which are not rare as the calculation depicted in Fig. 1 shows a clustering there. Although for more deformed nuclei the splitting between the two high energy components becomes smaller it should still be taken into account, at variance to what is often done. The increase of the width as predicted from Eq. (6) causes these two to have reduced height although all components have equal strength. Special care is needed for nuclei near closed shells as was argued for Fig. 1, the predicted *β* have to be reduced to *β*$_{eff}$ for nuclei near magic shells; the resulting predictions for the IVGDR curves then agree better to the data.

These curves result from TLO with an integrated cross section equivalent to the classical sum rule, divided equally between terms for k=1-3 [16, 18, 20, 62]. Thus the TLO-prediction for the absorption cross section into the isovector E1-mode $\sigma_{abs}^{E1,IV}(E_\gamma)$ is obtained by summing over three components k=1,2,3:

$$\sigma \cong 5.97 \cdot \frac{ZN}{A} \frac{2}{3\pi} \sum \frac{E_\gamma^2 \cdot \Gamma_k}{(E_k^2 - E_\gamma^2)^2 + E_\gamma^2 \Gamma_k^2} \text{ fm}^2 \quad (8)$$

Equivalence to the main term of Eq. (5) is obvious from the numerical normalization factor.

The deformation induced modifications of $E_k$ and the energy dependence of $\Gamma_k$ are seen in Eqs. (6) and (7), and agreement to experimental IVGDR data, depicted in Fig. 3, is reached by using only two global parameters, $m_{eff}$ and $c_w$. In fact, the latter one is not really free: As mentioned in view of Fig. 3, it corresponds to the $\Gamma_r$ for $^{208}$Pb, for which we get the same as predicted by a schematic calculation for the damping of dipole resonances formed from p-h states in a shell model [44]. Actually this calculation also justifies the use of a Lorentzian shape for the envelope over a very large number of close lying levels composing a doorway state. A different quantum aspect of our ansatz is related to the variance of the calculated deformation values (*cf.* Fig. 1) as also extracted from the HFB-calculations. In Fig. 3 not only the mean (*i.e.* expectation) values are used, but the quantal uncertainty is also depicted: Instantaneous shape sampling (ISS) was proposed [53] for this issue and we applied it as seen in Fig. 3; this has been already done for Mo isotopes [16].

The HFB-calculations [32] used for information on nuclear radii and deformation are available also for nuclei outside the valley of stability and this opens a possibility for a global prediction of nuclear photon strength also for heavy (A>60) exotic nuclei. The absorption cross section at low energy, important for radiative capture, has to be derived by extrapolation; the reduction of $\beta$ near magic shells as proposed here can be shown to have a minor influence for this. In any case the now applied predictions for the ground state triaxiality explains the IVGDR data without taking widths $\Gamma_k$ and energies $E_k$ from independent fits for each isotope. The agreement to the data for even Nd-isotopes to Eq. (8) is depicted in Fig. 3 and the 3 pole energies $E_k$ are indicated as black bars.

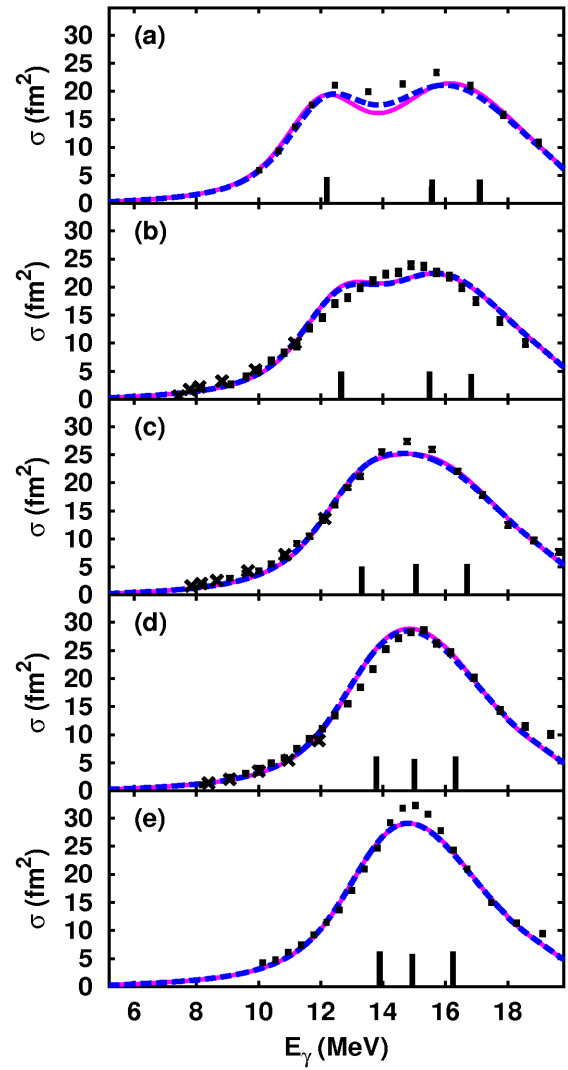

*Fig.3*: Cross section of photo-neutron production on the even $^{142}$Nd (e) to $^{150}$Nd (a) in comparison to the sum of three IVGDR-Lorentzians (TLO, dashed blue). The drawn magenta curves show the effect of shape sampling. Data of two experiments are overlaid: squares [63,■] or crosses [64,x], respectively.

The resulting parameterization [18, 19] had been tested thoroughly [16, 17, 18, 21, 24] on many experimental data available for the IVGDR widths, energies and strengths in various stable nuclei. In this paper we have added a prediction for the chain of Nd isotopes, for which new data [64] were published recently.

The interaction of heavy nuclei with photons of energy above the neutron separation energy $S_n$ is mostly resulting in reactions of type ($\gamma$,xn), and the experimental data points are electronically readable from a compilation [61]. The newly available quasi-monochromatic photon beams open unprecedented experimental possibilities. Several facts have to be regarded when judging the quality of older experimental photo-neutron data:

1. Considerable discrepancies were reported for experiments performed at different laboratories [65]. A reduction in the order of 10(4) % was found to be necessary for data obtained at Saclay [65, 66, 17, 67]; in our plots this was accounted for.

2. In a number of cases the ($\gamma$,p)-channel exhausts a portion of the photo-absorption cross section [60, 16, 13].

3. Most of the data were obtained by using quasi-monochromatic photon beams with a rather large energy resolution not much smaller than the predicted width of the IVGDR distribution. As it is often not well known experimentally we follow a proposal [66, 67] and assume it to be 0.6 MeV.

Items 2 and 3 influence the representation of the IVGDR peak region, but they do not have a significant effect on the tail a few $\Gamma$ below the pole where the contribution from the isovector E1 strength to radiative neutron capture is strong. Fig. 4 shows results for the Nd-chain; results of a similar analysis for the Mo's have been published previously [14, 16]. There it was shown, how the possible influence of all open channels on the extraction of the absorption cross section from the existing data can be tested by Hauser-Feshbach calculations and the code TALYS, which may have to be modified to incorporate the photon strength as derived by TLO. As the radiative neutron capture data discussed later are from the region of unresolved resonances, as are the photo-neutron data, it is indicated to use data averaged over photon energy to extract strength functions $f_\lambda$ from them; as shown [68] long ago, these can describe photon absorption as well as the electromagnetic decay.

## V. Photon strength in the IVGDR and below

The multipole strength functions $f_\lambda(E_\gamma)$ are related to the average photon absorption cross section in a given energy interval $\Delta E$ by:

$$f_\lambda(E_\gamma) = \frac{\langle \sigma_{abs}^\lambda(E_\gamma) \rangle}{g_{eff}(\pi \hbar c)^2 \overline{E}_\gamma^{2\lambda-1}}$$
$$= \frac{1}{g_{eff}(\pi \hbar c)^2 \overline{E}_\gamma^{2\lambda-1} \Delta E} \int_{\Delta E} \sigma_{abs}^\lambda dE. \quad (9)$$

The strength functions $f_\lambda(E_\gamma)$ are supposed to be direction independent and they are thus used for excitation as well as decay processes relating photon scattering to radiative capture and photonuclear processes [68]. Using that $f_\lambda$ is direction independent and thus also related to the electromagnetic decay widths of the resonant levels R in the interval $\Delta$ one gets:

$$f_\lambda(E_\gamma) = \frac{1}{\Delta E} \sum_{R \in \Delta E} \frac{\Gamma_{R\gamma}}{E_\gamma^{2\lambda+1}} \simeq \frac{\langle \Gamma_{R\gamma}(E_\gamma) \rangle}{D_R \overline{E}_\gamma^{2\lambda+1}}. \quad (10).$$

The quantum-mechanical weight factor $g_{eff}$ will be discussed below. $D_R$ denotes the average level spacing at the upper of the two levels connected by $E_\gamma = E_R - E_f$ and for constant $f_\lambda(E_\gamma)$ a decrease of the average resonance widths with increasing level density $\rho_R = 1/D_R$ is expected. Then a decay takes place between levels which are both excited and there is no simple way to study them starting from target ground states. But the average quantity $f_\lambda$ is insensitive to details of the nuclear spectrum and we approximate any electromagnetic transition strength of energy $E_\gamma$ by $f_\lambda(E_\gamma)$ to be independent of the energies $E_R$ and $E_f$; this assumption is called Axel-Brink hypothesis [69, 70].

For even nuclei with $J_0=0$ the $g_{eff}$ in Eq. (10) are identical to the quantum-mechanical weight factor as used in Eq. (3) with spins $J_0$ of the ground state and $J_r$ of the excited level. For $J_0 \neq 0$ the $f_\lambda$ used in Eqs. (9) and (10) are identical and this is based on two facts:

(a) Photon absorption into a mode λ populates m members of a multiplet with m=min(2λ+1, 2$J_0$+1). The observed strength corresponds to the cross section summed over the multiplet and this is described by an effective spin factor:

$$g_{eff} = \sum_{r=1,m} \frac{2J_r + 1}{2J_0 + 1} = 2\lambda + 1. \quad (11)$$

(b) The ground state widths $\Gamma_{0r}$ of each member of the multiplet are equal.

Both conditions were shown to be fulfilled in many heavy nuclei [68]; they follow from the assumption of weak coupling between the odd particle and the collective mode λ. In contrast to Eq. (9) there is no spin weight factor g in the numerator of the expressions Eq. (5) and (8) quantifying the sum rule and the IVGDR cross section. For zero ground state spin it is compensated by the factor 3 in the denominator [38]. In the case of scattering by a target with non-zero ground state spin $J_0$ the observed strength corresponds to the cross section summed over a multiplet as described with Eq. (11) and the statistical factor which would have to appear is 2λ+1; in such nuclei the IVGDR is a triplet corresponding to λ=1 (or a doublet for $J_0 = ½$).

The TLO-calculations for odd-A nuclei as shown in Figs. 4b and 4d were performed on the basis of these considerations and obviously they agree to the experimental data similarly well as is the case for even-even nuclei. In nonzero spin nuclei deformations and radii are from averaging the respective predictions [32] for the even neighbours; for near shell nuclei the deformation was reduced as for Figs. 1 and 3. A semi-microscopic HFB calculation [71] is also shown in Fig. 4; it assumes two resonance parts only with the same width. Each of them represents half of the total strength – a surprising choice in view of three oscillation axes, but apparently leads to a better fit to the data.

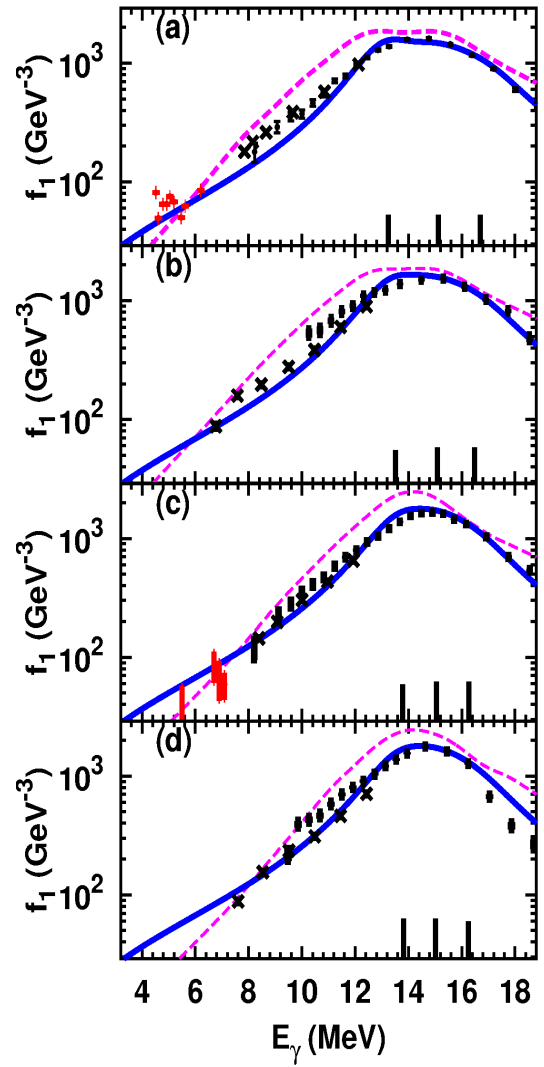

*Fig 4*: *TLO predictions (blue continuous curve) for the IVGDR in $^{146}$Nd (a) to $^{143}$Nd (d), in logarithmic scale with the pole energies shown in black. The measured cross sections of photo-neutron production are shown as squares [63,■] or crosses [64,x] respectively; their decrease at low energy may be a threshold effect. The*

*dashed magenta curves depict results of a different HFB-calculation [71, 72] assuming two poles only.*

## VI. Radiative neutron capture

The radiative capture of neutrons is of special interest for numerical simulations related to nuclear power equipment and for network calculations of astrophysical element production. The good agreement of the low IVGDR energy slopes to our 'triple Lorentzian' (TLO) using theoretical information on nuclear deformation including triaxiality [32] suggests the use the corresponding photon strength function also for other electromagnetic processes like radiative neutron capture. In the following discussion we present a schematic scheme with approximations: Above separated resonances Porter-Thomas fluctuations are reduced such that they can be neglected by averaging over a large number of neutron resonances r for which we assume $\Gamma_\gamma \ll \Gamma_n$. In a semi-classical approximation [73, 74] one gets for the capture cross section

$$\sigma_c(E_n) \equiv \langle \sigma(n,\gamma) \rangle_r \quad (12)$$
$$\cong 2\pi^2 \lambdabar_n^2 \sum_{J_b}(2\ell_n + 1) \int_0^{E_r} f_1(E_\gamma) E_\gamma^3 \rho(E_b, J_b) dE_\gamma$$

In this approach the neutron angular momentum $\ell \cdot \hbar = p_n \cdot R_A$ is calculated classically and any $\ell$-dependent neutron strength enhancement is neglected. In Eq. (11) the photon widths $\Gamma_\gamma$ are contained in the strength functions averaging in both intervals $\Delta R$ and $\Delta f$ for all resonances r and final states $f \in \Delta f = [0, S_n + E_r]$. For a first test of the "triple Lorentzian" strength function (TLO) for the case of radiative neutron capture the investigation can be limited to even-even target nuclei with spin 0. Then, the statistical factor, which accounts for the number of spin states reached by the γ-decay, may be set to 3 with sufficient accuracy. For such nuclei as studied by resonant neutron capture experiments [75, 72] the level density $\rho(S_n)$ is reasonably well known near $S_n$ with a mean accuracy of less than 20%. As the level density in the final nucleus enters strongly in Eq. (12) we have performed a critical review [76] of the studies predicting it in a Fermi gas picture.

Fig. 5 depicts the integrand in Eq. (12) versus the photon energy – based on the approximations listed – and thus shows the sensitivity of eventual predictions of radiative capture yields on the components of the dipole strength. Apparently it peaks as low as at ≈ 3 MeV; below, the factor $E_\gamma^{2\lambda+1}$ reduces the transition rates and above the density of levels to be reached becomes small. At this low energy the level density was estimated by a constant temperature model, such that the figure gives a schematic view on the situation only. The sensitivity of the radiative capture cross section against $f_1(E_\gamma)$ can be quantified by forming the ratio between the two dashed curves in Fig. 5. As was shown [76], the level density in heavy nuclei depends on their symmetry and an eventual breaking of it. This is assumed for the TLO-approach and this aspect will be discussed now.

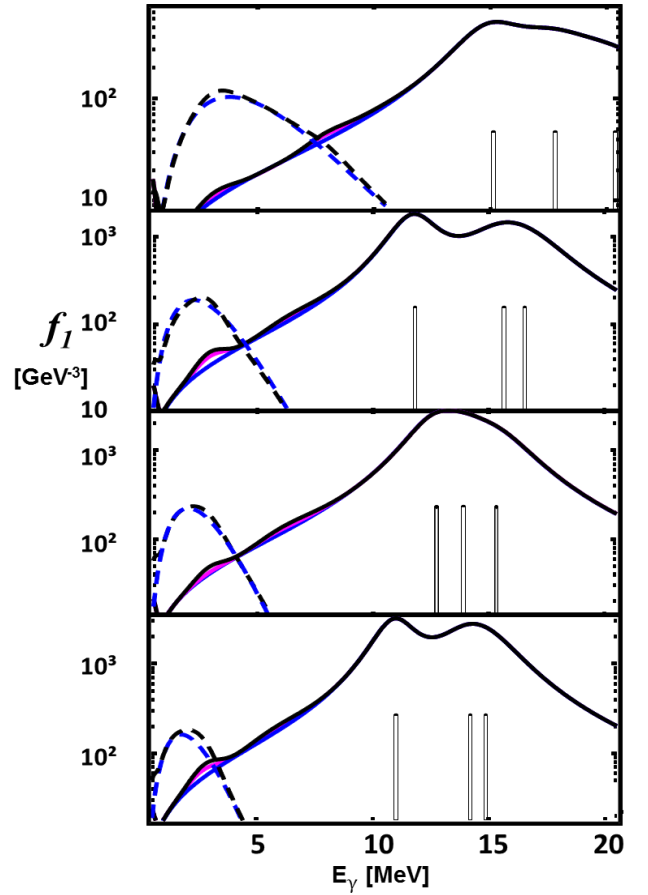

*Fig 5: For the 4 nuclei $^{78}$Se, $^{168}$Er, $^{196}$Pt and $^{240}$Pu (top*

*to bottom) the dipole strength f1(Eγ) in GeV⁻³ is depicted vs. the photon energy in MeV. The sum of three Lorentzians for the IVGDR (lowest blue curve) is shown together with the added contributions from minor (middle curve in magenta; [21]) dipole strength; the total sum appears in black as top curve. The dashed curves represent the sensitivity of the radiative capture against $f_{E1}(E_γ)$, (TLO, lower blue curve) and for the total $f_1(E_γ)$ (black top curve), cf. Eq. (12).*

## VII. Level densities and collective enhancement

### a. Intrinsic state density

Nuclear level densities $\rho(E_x, J^\pi)$ determine the final phase space for predictions of compound nuclear cross sections and decay rates. In a compilation [72] of data extracted from nuclear spectra and neutron capture resonance spacings it was shown that when they are parameterized in various ways, this may lead to inconsistent predictions. Hence the need of calculations based on fundamental principles, which rely on very few free parameters only, is indicated. Here a clear distinction has to be made between the intrinsic quasiparticle state $\omega_{qp}(E_x)$ and the level density $\rho(E_x,J)$ in the observer's system. An analytical approach often used for the calculation of $\omega_{qp}$ based on Fermi gas theory was proposed [77] to be better combined to a simple exponential dependence on energy for low excitation, where nuclei can no longer be considered a gas of Fermions. At variance to that work [77] we use the concept of a transition between a phase of nucleons paired to bosons and an unpaired gas. For atomic and molecular gases a 'critical' temperature $t_{pt} = \Delta_o \cdot e^C/\pi = 0.567 \cdot \Delta_0$ (with the Euler constant C=0.5772) was defined, which can also be applied to nuclei [78, 79, 80]. In the low energy ($E_x < E_{pt} = \tilde{a} \cdot t_{pt}^2 + E_{bs}$) regime we use

$$\omega_{qp}(E_x) = \omega_{qp}(0) \, exp\left(\frac{E_x}{T_{ct}}\right) \quad (13);$$

this corresponds to a constant temperature (CTM) model.

For energies above $E_{pt}$ the Fermi gas expression [81, 77] holds:

$$\omega_{qp}(E_x) = \frac{\sqrt{\pi} \cdot exp\left(2\sqrt{\tilde{a}(E_x - E_{bs})}\right)}{12 \, \tilde{a}^{1/4} (E_x - E_{bs})^{5/4}} \quad (14).$$

The parameter ã relates energy and temperature of a Fermi gas; it is often (confusingly) called level density parameter and even used as a variable to be fitted. We keep it fixed and derive the backshift energy $E_{bs}$ by subtracting the mass $M_{ld}$ given by a liquid drop formula from the measured mass $M_{exp}$:

$$E_{bs} = M_{exp} - M_{ld} + E_{co} \quad (15).$$

The backshift $E_{bs}$ represents the energy between the Fermi gas zero and the ground state of finite nuclei, and it corrects for the nuclear binding in shells. Expression (15) assures that shell effects, as well as pairing, are treated equally for even and odd nuclei in the Fermionic regime ($E_x \geq E_{pt}$). In $E_{bs}$ we include a pairing condensation term $E_{co} = \frac{3}{2\pi^2} \tilde{a} \Delta_0^2$ [80]. To avoid any fitting here we use an approximated pairing parameter $\Delta(E_x=0) = \Delta_0 = 12 \cdot A^{-1/2}$ and for ã we insert the nuclear matter value (with Fermi energy $\varepsilon_F = 37$ MeV):

$$\tilde{a} = \tilde{a}_{nm} = \frac{\pi^2 A}{4\varepsilon_F} \cong \frac{A}{15} \quad (16).$$

This approach [82] characterizes the Fermi gas by a gap $\Delta(t)$ falling with rising temperature parameter t down to 0 at a 'critical' temperature $t_{pt}$. The general features of this phase transition are evaluated by canonical thermodynamics for nuclear matter, and again we have no free fit parameter. It was shown [79] that the expression given by Eq. (14) for $\omega_{qp}$ in the Fermi gas regime – initially derived neglecting pairing [81, 77] – is a good approximation for the formalism derived with a micro-canonical inclusion of pairing, if $E_{pt}$ is back-shifted by the condensation energy $E_{co}$, which already appears in infinite Fermionic systems and which is independent of A. Unfortunately, that work [79] neglects the shell correction present in finite nuclei, but we find no significant differences in $\omega_{qp}(E_x \gtrsim S_n)$ when including it. As in earlier work [77] the energy dependence of the state density is assumed to be exponential for lower energies, *i.e.* in the pairing dominated phase below the phase transition point. This finds support in a recent analysis of level density data [77, 80, 83, 84]; we adjust $\omega_{qp}(0)$ as the state density at the lower end of the interpolation just above the ground state and it can be fixed there locally by regarding known spectral data ($E_x < \Delta_0$, $J$), similar as has been done previously [35, 85, 86]. We also tested a global approximation by setting $\omega_{qp}(0)=0.3/\Delta_o$ for the state density at the lower end of the interpolation region, and this has a minor effect for $E_x \gtrsim E_{pt}$, as in most nuclei $E_{pt}$ is smaller than the neutron binding energy $S_n$.

This is shown in Fig. 6, which also depicts the variation of $E_{bs}$ versus A, which is especially strong near closed shells. It also becomes obvious, that in the Fermi gas regime $E_{bs}$ is closely correlated to $E_{pt}$ and thus also to $\omega_{qp}(E_x)$; it is hence the quantity of higher importance for a level density prediction as $S_n$, at variance to a previous assumption [77].

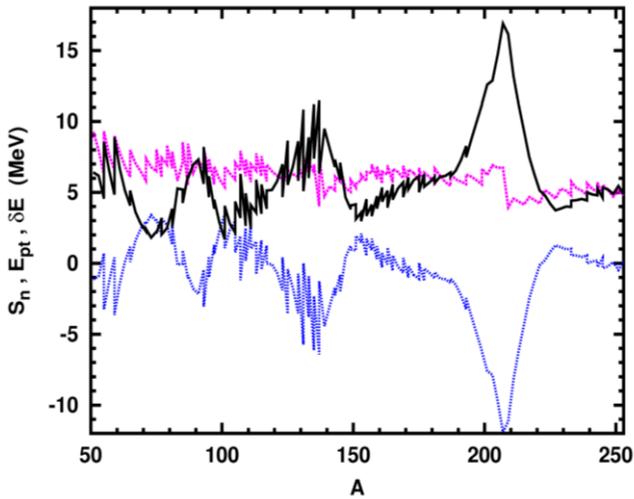

*Fig 6:* Phase transition energy $E_{pt}$ for nuclei in the valley of stability vs. A (full curve in black) in comparison to values for $S_n$ (dashed in red) and the shell correction energy $E_{bs}$ (lower curve in blue dots), all in MeV.

To demonstrate the energy dependence of the state density formalism presented here, results for $^{81}$Sr and $^{113}$Cd are given in Fig. 7.

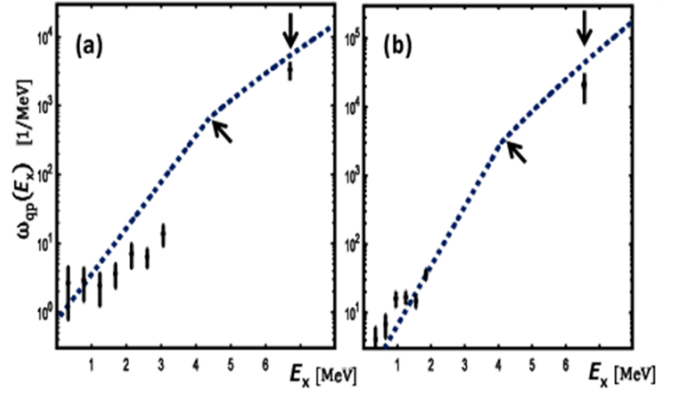

*Fig.7:* For the state density $\omega_{qp}(E_x)$ in the spin ½-nuclei $^{81}$Se and $^{113}$Cd the prediction is shown together with respective data from Ripl-3 [72] obtained using discrete levels (black ∎) [75] as well as resonance spacings (below vertical arrow); both are converted into state density $\omega$ by inverting Eq. (19). A change in slope at the phase transition energy $E_{pt}$ is clearly seen (diagonal arrow). The dotted curve in blue depicts Eqs. (13 to 16).

One sees in the figure, that an apparent nuclear temperature $T_{app} = \frac{\omega}{\partial\omega/\partial E}$, differs from $T_{ct}$ of Eq. (13) and this results in a change in the slope of $\omega(E_x)$ at $E_{pt}$. A slope change is well-known for the entropy at a 2$^{nd}$ order phase transition; in finite nuclei it is expected to be washed out, but in a rather old neutron scattering study [87] such a break was experimentally observed (in arbitrary units only – unfortunately). Close to magic nuclei the large (negative) shell correction results in a large break at the now large $E_{pt}$, but for weakly bound nuclei $E_{pt}$ becomes smaller. For completeness: the temperature as defined for a Fermi-gas $t = \sqrt{(E_x - E_{bs})/\tilde{a}}$, [77, 1, 79] is smaller than $T_{app}$ by up to 35%.

### b. Collective enhancement

The quantities to be compared to observed level spacings have to be derived from $\omega_{qp}(E_x)$ by a projection on fixed angular momentum J in the observer system. The proposal was made [81, 77, 88, 89] to consider the M-substate distribution of $\omega_{qp}(E_x)$ as Gaussian with width σ around M = 0 and to differentiate at M = J+½ with respect to M.

The redistribution of the quasi-particle M-states into levels of distinct spin J as incorporated here implicitly assumes [89] the nucleus to be exactly spherically symmetric. This assumption leads to a spin dependent level density [90, 1, 79, 91, 92, 93, 80, 72]:

$$\rho_{sph}(E_x, J) \cong \frac{2J+1}{\sqrt{8\pi}\,\sigma^3} e^{-\frac{(J+½)^2}{2\sigma^2}} \omega_{qp}(E_x)$$

$$\xrightarrow{\text{small J}} \frac{2J+1}{\sqrt{8\pi}\,\sigma^3} \omega_{qp}(E_x) \quad (17).$$

For the spin dispersion factor $\sigma^2$ a Thomas-Fermi approximation to the shell model predicts [94] a value $\sigma^2 \approx \sqrt{2A}$, which is smaller by nearly a factor of 3 as compared to the one from the 'statistical' moment of inertia $\mathfrak{I}_{st}$ with $\sigma^2 \cong \frac{\mathfrak{I}_{st}\cdot T}{\hbar^2}$, and $\mathfrak{I}_{st}$ assumed to be equal to the 'standard' rigid rotor value [5, 90] $\mathfrak{I}_{rig} = ⅖ \cdot M_A R_0^2 (1+⅓\beta)$ with $M_A$ and $R_0$ standing for nuclear mass and radius. Even when spherical symmetry was not assured for the nuclei studied, Eq. (17) has found widespread use [77, 95, 79, 91, 94, 80, 93, 72], often with $\mathfrak{I}_{rig}$ inserted.

Eq. (17) neglects strongly mixed collective modes which are pulled from their original quasi-particle energy down into the low excitation region. A proposal [82] to account for the broken spherical symmetry causing a large rotational collectivity yields a level density increase by a factor $\sigma^2$ (*i.e.* ≈ A/5) as compared to Eq. (17). This enhancement of level densities results from the build-up of a rotational band on each intrinsic quasi-particle state: *The total level spectrum, for a given angular momentum, is therefore obtained by summing over a set of intrinsic states rather than by a decomposition of the level spectrum, as for a spherical system* [90]. The resulting increase was included in some work on heavy nuclei [88, 89, 96] by an additional term for 'collective enhancement'. But still an agreement with observations was not reached without a significant enlargement of the 'level density parameter' [72, 86] as compared to the nuclear matter value ã. It was eventually adjusted in a fit, a method which we consider highly questionable. In addition, an excitation energy dependence of ρ was introduced [91] − at variance to the Fermi gas, which we accept as proper description of the statistics in highly excited nuclei.

In view of the quite common triaxiality [32] and the related 3-fold splitting observed in the IVGDR for nuclei with A>50 – as described previously [18, 19] and in sections IV and V – we disregard case a. Instead we studied the effect of allowing the breaking of various symmetries in the Fermi gas regime and at $E_{pt}$. Especially the absolute value of predicted level densities is expected to increase due to additional degrees of freedom. Surprisingly, this topic was rarely [97] taken up by subsequent studies, most of which only regarded axially deformed or spherical nuclei. Actually, it can be generalized to even allow the breaking of any symmetry; in the limit of low J, large σ and negligible $E_{yr}$ one obtains [90] approximate formulae for one parity:

a. spherical case: $\boldsymbol{\rho}(E_x, J^\pi) \rightarrow \frac{2J+1}{2\cdot\sqrt{8\pi}\,\sigma^3} \omega_{qp}(E_x)$

b. axial symmetry $\Rightarrow \boldsymbol{\rho} \rightarrow \frac{2J+1}{2\cdot\sqrt{8\pi}\,\sigma} \omega_{qp}(E_x)$ (18)

c. non-axial (triaxial) $\Rightarrow \boldsymbol{\rho} \rightarrow \frac{2J+1}{2\cdot 4} \omega_{qp}(E_x)$

d. no reflection symmetry $\Rightarrow \boldsymbol{\rho} \rightarrow \frac{2J+1}{2} \omega_{qp}(E_x)$

By the transition from case c to case b (nuclear body is symmetric with respect to one axis) a decrease of the level density by $\sqrt{\pi/2}\cdot\sigma \gtrsim 4$ (for A≈160) is expected in the limit of small J; a reduction by $\sqrt{\pi/2}\cdot\sigma^3 \gtrsim 80$ is the result of a change from case c to case a (Eq. (18c) to Eq. (17)), valid for the level density of completely spherical nuclei [82, 90, 1]. The size of these factors indicates that the dependence of the absolute level density on the symmetry of the nuclei is appreciable, whereas in cases c and d the size of the deformation does not enter in the low J limit. This limit is important as it is the case of levels populated in the capture of slow neutrons by even nuclei, which is the source of accurate level density data at $S_n$. It is obvious from Eqs. (17 to 19 and 21) that the deformation parameters $\beta$ and $\gamma$ only effect $\rho(E_x, J^\pi)$ by the spin cut-off term, whereas the symmetry class is of greater importance.

The inclusion of broken axial shape symmetry is considered in the present study and Eq. (18c) follows a proposal made long ago [90]: For an equilibrium shape *that possesses all the rotational degrees of freedom of a three-dimensional body* and thus *completely violates rotational symmetry, in the sense that it is not invariant with respect to any rotation of the coordinate axes*, such a rotational band on top of every intrinsic state *involves (2J+1) levels with total angular momentum J. Each of these levels is itself (2J+1)-fold degenerate, corresponding to the different components M.*

For broken spherical symmetry ($R_{1,2} \neq R_3$) a collective rotation becomes possible. Then (and also for $R_1 \neq R_2$) any rotation results in an yrast band with $E_{yr}(I)$ and no levels with spin I exist below this energy. In the extension of Eqs.(18 b-d) to larger spin values a cut-off is induced similar to the one in Eq. (17), but it now results from this yrast energy $E_{yr}(I)$. The yrast state is quasi the ground state for all levels with spin and parity $I^\pi$ formed by combining the collective angular momentum I and the intrinsic M-state distribution like in Eq. (17). Assuming that the excitation energy $E_x$ is large as compared to the collective energy $E_{yr}(I)$ one gets [82, 77, 90]:

$$\rho(E_x, I^\pi) = \frac{2I+1}{2 \cdot 4} \omega_{qp}(E_x - E_{yr}(I))$$
$$\approx \frac{2I+1}{2 \cdot 4} \omega_{qp}(E_x) \cdot e^{-\frac{E_{yr}(I)}{T_{eff}}} \quad (19).$$

Here one approximates the energy dependence of $\omega_{qp}(E_x)$ by the constant temperature formula from Eq.(13) with $T_{ct} = T_{eff}$. Compared to Eq. (17) an increase in $\rho(E_x, I^\pi)$ is found, which was already seen in Eq. (18c). The resulting astonishingly very simple expression for small J given there is also given in the book by Bohr and Mottelson [1, *cf.* Eq. (4-65b)]. The numerical factors 2 and 4 are related to parity conservation and to the invariance with respect to rotations by 180° about any axis. The latter should hold for quadrupole interactions [1], but may possibly be broken in the presence of octupole deformation.

As pointed out earlier [76], various assumptions are made and they are listed here albeit not all of them have a large influence for our conclusions:

1. Quasi-particle states are evenly spaced (at least on average) at the Fermi energy, not varying with neutron excess N−Z.

2. The pairing parameter $\Delta(E_x=0)$ is approximated by $\Delta_0=12 \cdot A^{-1/2}$, independent of J and N−Z.

3. For the control of ã the Fermi energy is taken to be independent of N−Z, $\varepsilon_F = 37$ MeV.

4. The influence of shell effects is controlled by $E_{bs}(Z,A)$, found by subtraction of the experimental mass from liquid drop values.

5. At variance to earlier work [91, 72, 80, 93, 94] the shell correction is directly applied to the backshift energy $E_{bs}$ [77, 95, 96].

6. A certain ambiguity concerns the back shift energy $E_{bs} = M_{exp} - M_{ld} + E_{co}$ and hence the liquid drop model parameters used for the calculation of $M_{ld}$. We concentrate on a proposal [97] which accounts for the shell effect on masses explicitly. It also regards deformation effects and obtains a good fit for ground state masses. It does not treat the breaking of axial symmetry, but its influence on ground state masses was calculated to be very small for most heavy nuclei [48, 98].

Replacing our favourite choice by the one from [94] increases the level density for actinide nuclei by nearly an order of magnitude, whereas the use of ref. [48] has the opposite effect. A recent liquid drop model fit to masses [99] based on fitting a volume and a surface term independently without a shell correction term delivers nearly equal ρ($S_n$, $J^π$) as compared to our choice, but another new liquid drop model (LDM) fit including a curvature term [100] leads to a significant over-prediction for actinide nuclei.

**VIII: Level densities for arbitrary spins**

Another point needing regard is the determination of $\mathfrak{I}$. Here an example for a very nearly axial nucleus is instructive: In $^{238}$U the yrast band is well described only above spin 20 by the standard value for $\mathfrak{I}_{rig}$, whereas near the ground state the level scheme indicates energies to be higher by ≈ 60% [101]. To get the spin integrated level density ρ($E_{lab}$) many spins have to be summed and thus we discuss in the following the influence of spin and we concentrate on the triaxial situation, Eq. (18c). To evaluate ρ($E_{lab}$,J) for J≫0 two facts will be discussed separately:

1. In the case of an intrinsic ground state spin $J_0$ the vector equation $\hat{J} = \hat{I} + \hat{j}$ leads to $E_i = E_x - E_{yr}$ and in Eq. (19) a replacement is indicated:

$$ρ(E_x, J^π) = m · ρ(E_i, I^π) \qquad (20).$$

Here m=min (2I+1, 2j+1) represents a weak coupling similar as leading to Eq. (11).

2. Levels containing an elevated collective angular momentum I move out of the reference region and this is accounted for by the generalized cut-off factor $e^{-\frac{E_{yr}}{T_{eff}}}$ already used in Eq. (19). Here the spin dependence of the yrast energy $E_{yr}$ on angular momentum has to be known. For the axial case the usual rotational expression $E_{yr}(I) = \frac{\hbar^2 · I(I+1)}{2 · \mathfrak{I}}$ is fine, but for broken axial symmetry we propose to replace I·(I+1) by I+c·I². Axiality is represented by c=1 whereas c=0 leads to a linear increase of $E_{yr}$ with I, similar to nuclei with large γ [33, 32]. A look at yrast level energies in the tables attached to the CHFB work [32] we have used for the IVGDR analysis suggests to identify c with cos(3γ). This leads to a modification of Eq. (19) and $E_{yr}$ is now approximated by:

$$E_{yr} = \frac{\hbar^2 · (I+c·I^2)}{2 · \mathfrak{I}_{app}}; \quad c = \cos(3γ) \qquad (21a).$$

From absolute excitation energies in the above-mentioned tables [32] we derive a rough estimate for the moments of inertia in dependence of γ and a reduction of $\mathfrak{I}_{app}$ to

$$\mathfrak{I}_{app} = \frac{1+64·c^2}{65} · 50/MeV \qquad (21b)$$

is suggested; the numbers 64/65 and 50 should be considered a first estimate and may well be changed in further study. Our very simplified schematic attempt to cover most heavy nuclei is much less sophisticated than a recent paper on the spin distribution of nuclear levels [102]. Like our work that study uses the framework of the spin cut-off model, but it is limited to A<60 and it does not consider the breaking of axiality. In our approach the reference to the yrast level with spin 4 is made in view of its presentation in the CHFB tables [32] and as important component in the sum of Eq. (22), which is centred at I= 4−6. The change by more than 20 from the limit c=0 to the other extreme c=1 is drastic, but the table reaches from quasi spherical triaxial to strongly deformed nuclei. Thus we set $\mathfrak{I}_{app}$ to be not just related to rotation but it parametrises smoothly the transition from a nucleus with a high energy 1$^{st}$ excited state to one with a rotation like band starting from a very low 2$^+$ state. In certain sense this gradual change is an analogue to the blue curve in Fig. 1 depicting the transition from nuclei with near zero $Q_i$ and cos3γ to those with large quadrupole deformation and axial symmetry. The deformation dependence of $\mathfrak{I}_{app}$ resembles the one of $\mathfrak{I}_{irrot}$ [5], but it approaches $\mathfrak{I}_{rig}$ in the axial limit c→1, as assumed in previous work [90].

The modifications used in Eq, (19-21) originate from axial symmetry breaking and result in an identical low spin limit like in Eqs. (18). The sum over spin I leading to the total (spin integrated) level density ρ(E$_x$) is likely to be also influenced by it. We could show by numerical tests with Eq. (21) and the new choice for $\mathfrak{I}_{app}$ and E$_{yr}$ that the subsequent Eq. (22), which is nearly identical to the proposal made previously [90], holds for this 'true triaxial' case with $\mathfrak{I}_{app}$:

$$\rho(E_x) = \sum_{I=0}^{\infty} \rho(E_x, I) \qquad (22a)$$

$$\frac{\rho(Ex)}{\omega_{qp}(E_x)} \cong \frac{3}{4} \cdot \sum_{I=0}^{\infty} (2I+1) \cdot e^{-\frac{E_{yr}(I)}{T_{eff}}} \qquad (22b)$$

$$\approx \frac{3 \cdot T_{eff}}{1 \text{MeV}} \sqrt{\frac{\pi \cdot \mathfrak{I}_{app}}{1 \text{eV} \cdot \hbar^2}} \approx 3 \cdot \sigma_{sco} \sqrt{\pi T_{eff}} \qquad (22c)$$

In variance to previous work [90] we replace in Eq. (22b) a 3-fold product by a factor 3 as account for triaxiality; our estimation for this equation by Eq. (22c) is accurate within 15% (cf. Table I). It shows an increase of ρ(E$_x$) over ω$_{qp}$(E$_x$) growing from ≈4 up to more than 15 when going from c=0 to c=1, *i.e.* for nuclei with small γ and large Q$_i$. For this 'axial' limit the increase of ρ(E$_x$) vs. ρ$_{sph}$(E$_x$) is still there, but smaller than for ρ(E$_x$,I), as apparent from comparing lines b and a in Eq. (18). This difference is the result of our approximate account for the change with triaxiality in Eq. (21), close to what is observed in collective nuclear excitations.

Here the now proposed decrease of $\mathfrak{I}_{app}$ with γ and the lower slope in E$_{yr}$(I) in Eq. (21) play an important role via the exponential spin cut-off. The increase of ρ(E$_x$, J$^\pi$) due to the breaking of axial symmetry is included in Eq. (22) and the ratio $\frac{\rho(Ex)}{\omega_{qp}(E_x)}$ is a measure of a collective enhancement. Our way to estimate it results in an approximately linear dependence on T$_{eff}$ and $\sqrt{\mathfrak{I}}$; instead of adjusting ã by a fit we use the nuclear matter value ã$_{nm}$. Table I shows the resulting γ-dependence of the estimations made in Eqs. (21) and (22). Compared to a previous estimate of 2-3 for this ratio ([80], including a rotational enhancement) we predict a clearly larger enhancement due to the increase in the number of degrees of freedom.

| γ [deg] | E(4+) [MeV] | $\frac{\mathfrak{I}_{app}}{\hbar^2}$ [1/MeV] | $\frac{\rho(Ex)}{\omega_{qp}(E_x)}$ (22b) | $\sqrt{\frac{\mathfrak{I}_{app} \cdot T}{\hbar^2}}$ (22c) | σ$_{sco}$ |
|---|---|---|---|---|---|
| 30.00 | 2.56 | 0.78 | 1.09 | 2.35 | 0.62 |
| 28.09 | 2.19 | 1.28 | 1.42 | 3.01 | 0.80 |
| 26.15 | 1.29 | 2.78 | 2.72 | 4.43 | 1.18 |
| 24.18 | 0.83 | 5.28 | 4.56 | 6.11 | 1.62 |
| 22.14 | 0.59 | 8.78 | 6.60 | 7.88 | 2.10 |
| 20.00 | 0.45 | 13.28 | 8.70 | 9.69 | 2.58 |
| 17.71 | 0.36 | 18.78 | 10.81 | 11.52 | 3.06 |
| 15.19 | 0.30 | 25.28 | 12.92 | 13.37 | 3.56 |
| 12.29 | 0.26 | 32.78 | 15.01 | 15.22 | 4.05 |
| 8.61 | 0.22 | 41.28 | 17.10 | 17.08 | 4.54 |
| 0.00 | 0.20 | 50.00 | 19.17 | 18.95 | 5.04 |

*Table I: Collective enhancement in dependence of triaxiality γ resulting from the approximations in Eqs. (21-22) and corresponding to T$_{eff}$=500 keV.*

This is also true for the last line representing large axial deformation and a rotational yrast line. But it was shown [11] that even for large Q$_i$ a small triaxiality γ is observed experimentally – in accordance to theory [33, 32]. The 6$^{th}$ column indicates a strong variation of the spin cut off parameter σ$_{sco}$ and we compare our ansatz to experimental data.

### IX. Comparison to experimental data

### a. Level densities

At first, a comparison will be presented for the energy region near the neutron separation energy S$_n$, for which good data are available from compound resonances for neutron capture in the eV and keV range. If even target nuclei are used and the neutron energy is low enough, only levels with spin ½$^+$ are observed and average level distances D(E, J$^\pi$)=1/ρ(E, J$^\pi$) deliver level density information. One may thus use Eq. (18c) if no extra proof of axial symmetry conservation exists; triaxiality is the more general assumption and thus needs no confirmation. For 132 nuclei with A>70 the average distance of respective s-wave neutron capture resonances is available [75]. As for spin ½ the small J limit in Eq. (18) is lower by a few % only as compared to the full expression, it is interesting to compare experimental data to this limit, as was done in Fig. 7.

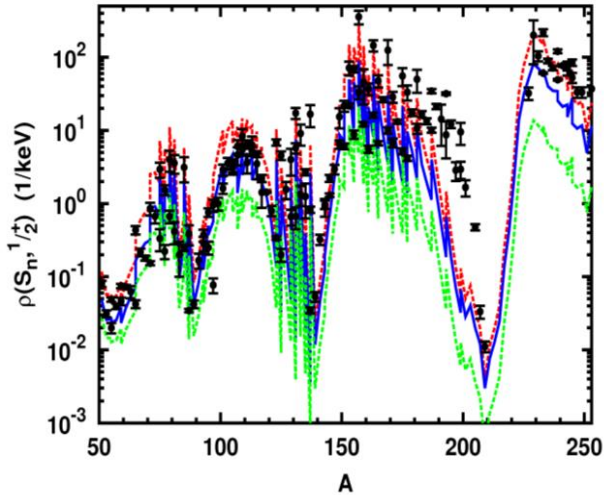

*Fig. 8*: *Average level densities ($S_n$, ½$^+$) in nuclei with 51<A<253 as observed in neutron capture by spin 0 target nuclei. Data (I) were compiled in RIPL3 [75, 72]; they are compared to our parameter-free prediction with an effective shell correction from a liquid drop calculation [97]. The lowest line (dashed green) corresponds to assumed axiality, whereas the drawn blue line depicts the triaxial calculation. A possible increase of ρ may result from an increased ã due to surface effects (red dashed line) discussed [76] previously.*

Fig. 8 depicts the results from using Eq. (18b&c) to measurements available for 132 nuclei and obviously many of the experimental points lie close to the prediction. This accord over many orders of magnitude on absolute scale is reached without any fitting, if the idea of axial symmetry is given up for heavy nuclei in the valley of stability. Our prediction is based on a widely used LDM fit [97] and it yields reasonable agreement to experimental level densities near $S_n$ and this may well be regarded as an additional indicator for its quality. Applying a damping of the numerator in Eq. (14), as proposed previously [95], leads to an improved agreement for A~200, but reduces the prediction near A~100. The also proposed rise of ã by 25% to account for surface effects leads to a small increase only, as shown in the figure. To significantly increase ρ(E,I) a much stronger rise of ã is needed, as has been applied in the past [80, 72] to compensate the ignored triaxiality.

Another experimental information [80, 72] on ρ(Z,N,$E_x$,$J^π$) stems from ensembles of discrete levels with equal spin populated in nuclear reactions at low $E_x$ and counting them up to an $E_x$, above which completeness is no longer assured. To test our approach for odd-n nuclei we selected 2 spin ½-isotopes with a satisfactory number of levels [72]. From counting bound levels, as well as resonances just above $S_n$, plots of ω($E_x$,J) were produced and depicted in Fig. 7. As their spins are known, only 2 obvious assumptions are needed to obtain $ω_{qp}(E_x)$ from the data on ρ($E_x$,J) by using Eqs. (19-20):

1. Parities are equally distributed.
2. The spin cut off factor $σ_{sco}$ in the exponential term can be taken from systematics [93].

This procedure of applying the spin dependent factors to the data allows energy differences between states of different spin to be shown in the same plot. But also spin integrated level densities can be regarded with respect to our predictions as made in Eqs. (20-22). They may be obtained in compound nuclear reactions from observing the yields of decay gamma rays from properly defined excitation regions. In experiments at the Oslo cyclotron the energy of one ejectile from a binary reaction is determined magnetically and the decay pattern is disentangled by a multi-detector device. The dependence of ρ(E) on the excitation energy is covered in small steps, but the absolute normalization has to come from other sources. At the low end the density of discrete levels is used in a way similar as we fix the CTM regime at the low energy. One example of special interest is shown in Fig. 9: Two well deformed nuclei, both with mass 238, but different ground state spins 2$^+$ and 0$^+$ are compared to each other and to predictions.

As their $E_{bs}$ are different, also their phase transition energies are so (3.8 and 4.6 MeV) and their intrinsic state densities $\omega_{qp}(E_x)$. And their level density $\rho(E_x)$ is collectively enhanced by ≈16 for both (see Table I) and by the additional spin factor 2j+1=5 from Eq. (20) for $^{238}$Np. In sum they are apart by a factor of 17±3, a value similar to the experimental finding.

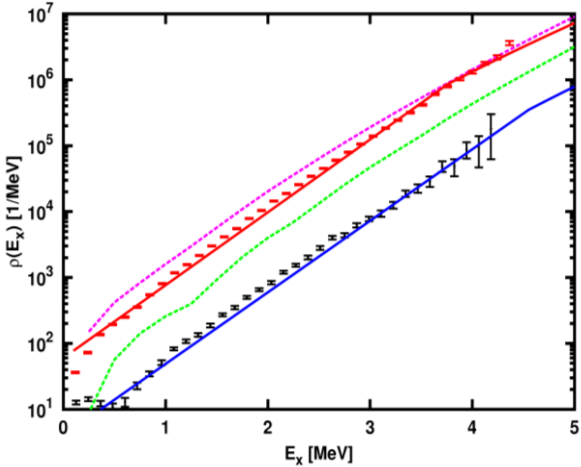

*Fig. 9*: Level densities ($E_x$) in $^{238}$U(lower data) and $^{238}$Np(top) as observed in the reactions $^{238}$U(d,dxγ) [103] and $^{237}$Np(d,pxγ) [104]. Data (**I**, **I**) are compared to our parameter-free prediction with an effective shell correction from a liquid drop calculation [97]. The full red curve depicts the calculation for $^{238}$Np whereas the blue full curve corresponds to $^{238}$U. Results of respective HFB calculations [105], summed for both parities, are depicted as dotted curves in magenta and green, respectively.

Hence the good agreement of our prediction to both data sets is remarkable and this finds support by less agreement with the HFB calculations [105], which are taken from the RIPL-3 project [72], listed as observables $\rho_{obs}$.

A similar comparison will now be presented for a completely different type of nucleus: For $^{92}$Mo the TLO analysis of IVGDR-data indicated triaxiality with γ≅30° [16] in accord to CHFB [32] and this is supported by the agreement shown in Fig. 10. Above $E_x$ = 2.5 MeV our prediction for the collective enhancement of 3.6 (cf. Table I) is close to the observation of the Oslo group [106, labelled 'rec' in the webpage], which do not reach the phase transition energy of 8.7 MeV, as predicted by us. The less convincing agreement with respective HFB predictions [go08, ca09] indicates a need to further improve such calculations, e.g. by application of the generator coordinate method (GCM) to have the proper angular momentum projection as required by quantum mechanics.

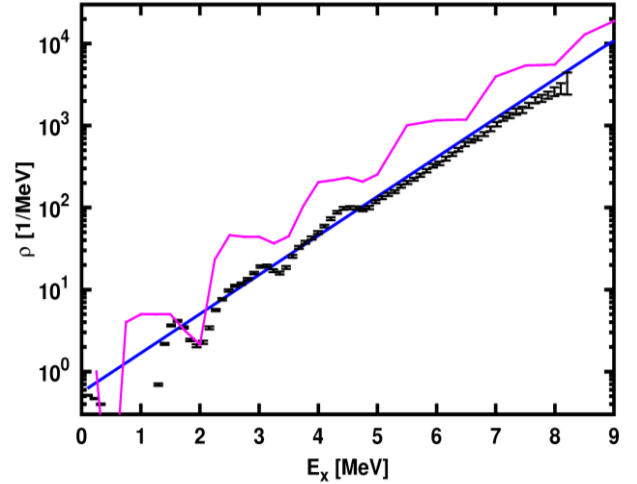

*Fig. 10*: Level densities ($E_x$) in $^{92}$Mo; experimental data [**I**, tv16] are compared to our parameter-free prediction depicted by the blue line. Results of respective HFB calculations [105, 72] are shown as the wiggly red curve.

In our theoretical reference [32] GCM is included; here we only use the broken axial symmetry as indicated by Eq. (18c). To quantify the estimations made in Eqs. (21 and 22) theoretical information on triaxiality and on $4^+_{yrast}$ was considered; the latter is also given in the supplementary material.

**b. Average radiative widths**

It was pointed out previously [107] that strength information can be extracted from capture data directly by regarding average photon widths $\bar{\Gamma}_\gamma$. These are proportional to the ratio between the level densities at the capturing resonances $r$ – included in $f_1(E_\gamma)$, as already used in Eq. (12) – and at the final states $b$ below $S_n$ reached by $E_\gamma = E_r - E_b$, and depend in addition on the photon strength in the low energy tail eventually extrapolated from the IVGDR. It is known that $\bar{\Gamma}_\gamma$ does not vary with $E_r$ [107, 68] and hence it can be approximated for $J_r = 1/2^+$ by summing over all final bound levels $b \in \Delta_b$, i.e. over $\Delta_b = [0, S_n + E_r]$:

$$\bar{\Gamma}_\gamma = \sum_{b \in \Delta_b} \Gamma_\gamma^{r \to b} \cong \rho(E_b, J_b) \cdot \langle \Gamma_\gamma^{r \to b} \rangle \cdot \Delta_b$$
$$\cong \int_{\Delta_b} \frac{\rho(E_b, J_b)}{\rho(E_r, J_r)} f_1(E_\gamma) E_\gamma^3 \, dE_\gamma \qquad (24).$$

Average radiative widths were derived by a resonance analysis of neutron data taken just above $S_n$ and tabulated [75] for 115 even-odd nuclei with $51 \leq A \leq 253$. These average widths $\bar{\Gamma}_\gamma$ allow a combined test of predictions for photon strength and level densities, and respective data are shown in Fig. 11.

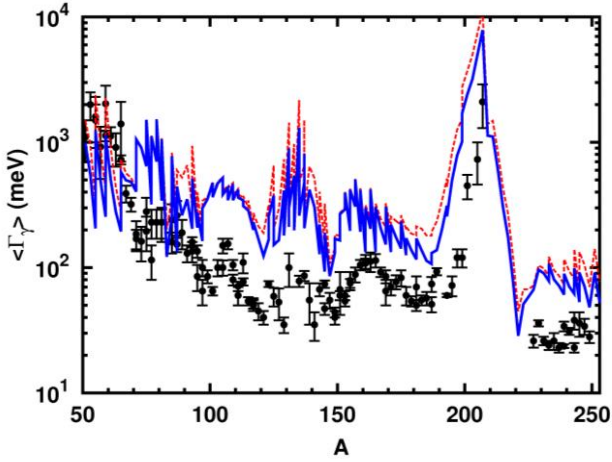

*Fig. 11: Average radiative widths as determined experimentally for nuclei populated by radiative neutron capture by spin 0 targets with A>50 shown in black [■, 75] are compared to calculations. The combination of the complete dipole strength described in sections V to VIII with the level density as given in section IX is depicted as drawn blue curve and the red dashes indicate an increase of $\omega_{qp}(0)$ by 3.*

Radiative capture into spin-0 targets through the s-channel ($\ell_n = 0$) is considered only, such that the spin cut off in Eq. (19) can be neglected and the approximation in Eq. (18c) is valid. As shown in Figs. 11 and 12, the agreement between prediction and data is satisfactory and we identify the following ingredients to be important:

1. A shell correction based on a LDM [97],

2. TLO with deformation values from HFB [32],

3. level densities enhanced by broken axiality.

There are local discrepancies – especially just below $^{208}$Pb – probably related to the neglect of shell effects (other than $E_{bs}(Z,A)$) in the proposed ansatz for the level density and the apparent disagreement for A < 70 may have a similar origin. The comparison to existing experimental data for $\langle \Gamma_\gamma \rangle$ as depicted in Fig. 11 uses the photon strength as discussed in sections III to V and the level density obtained from the parameterization described in section VII. This is at variance to previous work which only covered limited numbers of nuclides and used parameters for $\rho(E_x)$ locally adjusted [107, 108, 109, 110, 111]. The good accordance to experimental data on absolute scale as shown in Fig. 11 enables an evaluation of the importance of various approximations applied: A decrease of $\omega_{qp}(0)$ by a factor of 3 modifies $\langle \Gamma_\gamma \rangle$ by 20 to 50 % when regarding nuclei with A ≈ 70 resp. A ≈ 240 and this shows that the agreement may be improved by an introduction of local information, not included in our prediction, which is explicitly based on global properties only.

A large effect is expected from the coupled mode $(2^+ * 3^-)_{1^-}$, studied theoretically [6] since long, but for its strength only scarce data scattered in A and Z are available [112], which has been used as a guide here. The increase of $\langle \Gamma_\gamma \rangle$ by magnetic dipole strength was indicated [109], but in a recent review [113] this strength was demonstrated to be significantly smaller. As shown there it is concentrated in isoscalar and isovector components of a giant magnetic resonance expected near $S_n$ and thus outside of the overlap peak in Fig. 5. Also the magnetic strength of a scissors mode is discussed there being much closer to this peak and of some importance for an enhancement of $\langle \Gamma_\gamma \rangle$. But still, it appears that, similar to non-nuclear systems, electric dipole modes dominate radiative nuclear processes. The increase due to the inclusion of minor strength [21] has been estimated for an average over A to increase $\langle \Gamma_\gamma \rangle$ by less than 1.5.

### c. Average radiative capture cross sections

As pointed out [114], the folding of experimental neutron capture cross sections as well as those given by Eq. (12) with a Maxwellian distribution of neutron energies is straightforward. In view of the fact that $D \gg \Gamma_r \gtrsim \Gamma_{r\gamma}$ the Maxwellian averages around 30 keV are formed incoherently with neglect of Porter-Thomas fluctuations:

$$\langle \sigma(n,\gamma) \rangle_{kT} \cong \frac{2}{\sqrt{\pi}} \frac{\int_0^\infty \sigma_c(E_n) E_n \cdot e^{-E_n/kT} dE_n}{\int_0^\infty E_n \cdot e^{-E_n/kT} dE_n} \quad (25).$$

By only regarding the radiative capture by spin-zero targets effects related to ambiguities of spin cut-off or dispersion parameters and angular momentum coupling are suppressed, but still the data vary by about 4 orders of magnitude in the discussed range of A – and are well represented by the TLO-parameterization used here together with the proposed ansatz for $\rho(A, J^\pi, E_x)$, as is obvious from Fig. 12. The overall agreement on absolute scale and over more than three decades is remarkable; a discrepancy observed in the region of A > 230 may well be related to an over-prediction low energy components in state density or strength function, which have a large importance for high nuclear masses.

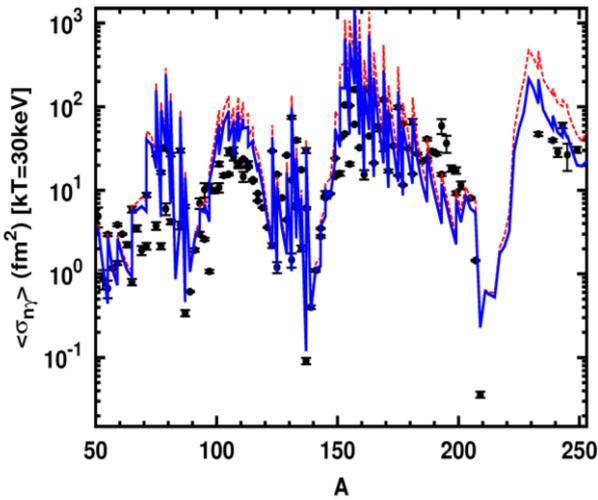

*Fig. 12: Maxwellian averages of measured cross sections for radiative neutron capture into even nuclei with J=0 and 50<A<250 (+,[115, 116]) for $kT_{AGB}$ = 30 keV. They are plotted vs. $A_{CN}$ in comparison to calculations based on Eq. (25) with TLO (dotted curve) and including the minor components (full curve). The level densities are determined as given in section VII.*

This and other local effects originating from details of the shell structure cannot all be treated in this paper, the main topic of which is the importance of triaxiality in heavy nuclei.

### X. Summary and outlook

Admission of axial symmetry breaking of heavy nuclei improves a global description of Giant Dipole Resonance (IVGDR) shapes by a triple Lorentzian (TLO). When theoretical predictions are used for the A-dependence of pole energies from droplet model [47] and spreading widths based on one-body dissipation [58], the TRK sum rule is obeyed quite well [14, 18]. Our new analysis of data for Nd isotopes based on a prediction for their triaxial deformation [32] demonstrates this also for cases with rather small splitting of the IVGDR. These were previously assumed to indicate a large spreading width and this makes an extension to energies outside of the IVGDR questionable.

Another effect – hitherto not emphasised as such – also indicates a breaking of axial symmetry in nearly all heavy nuclei: Without any fit of parameters the scheme based on non-axiality reproduces observations for level densities in nuclei with A > 50 surprisingly well on an absolute scale. Here the Fermi gas prescription is only used above a phase transition at the critical temperature, the pairing condensation energy is included in the backshift and the collective enhancement due to symmetry breaking is included. Some influence on the intrinsic state density $\omega_{qp}(E_x)$ as well as on the level density $\rho(E_x)$ was found to emerge from the choice made for $E_{bs}$, by which the Fermi gas zero is fixed with respect to the nucleus' ground state. Here an uncertainty arises from the various LDM fits to ground state masses and this, together with an ambiguity in the moments of inertia indicates a need for comparisons of our fit-parameter free ansatz to experimental data. We present here some such comparisons showing good agreement for the choice proposed by us.

This leads to the conclusion that the combination of the TLO-based photon strength approach to the 'triaxial' Fermi gas formalism for level densities predicts neutron capture in the range of unresolved resonances – including Maxwellian average cross sections compiled recently [115] for $\langle E_n \rangle$=30 keV – reasonably well. A good understanding of the radiative capture of fast neutrons is an important part for any theoretical attempt in the direction of the transmutation of nuclear reactor waste [22, 23, 24]. We predict level densities in actinides well and we showed that TLO produces a good description of IVGDR data for $^{197}$Au [15]; our approach may be a guideline for predictions for radiative capture of fast neutrons by actinide nuclei.

Five points are made here again:
1. The triple Lorentzian (TLO) description of IVGDRs and their low energy tail agree well to data without any modification by extra energy dependence. Deformation and triaxiality values are taken from CHFB-calculations [32] with the generator coordinate method assuring good angular momenta by a proper spin projection.

2. A direct account for broken symmetries leads to a good agreement of the TLO-predictions to experimental IVGDR shapes in accord to the TRK sum rule without any fit parameters other than an effective mass for resonance energies and one parameter to fix their widths versus $E_x$; both are global for all heavy nuclei studied [19].

3. By admitting broken axial symmetry the rather common procedure to predict level densities is modified such that agreement to data on absolute scale is obtained without an adjusted 'level density parameter' and ad hoc assumptions on collective enhancement made previously [72].

4. The predictions for average radiative widths and Maxwellian cross sections are sensitive to three not fully controlled ingredients: The state density $\omega_{qp}(0)$ near the ground state of the final nucleus, the moments of inertia and the extraction of the back-shift energy from liquid drop model fits.

5. Low energy dipole strength induced by modes other than the IVGDR has some influence on capture yields [21], and further experimental studies would help to better quantify this.

### XI. Conclusions

The breaking of axial symmetry in excited heavy nuclei is an important feature for the analysis of giant dipole resonance data presented here once more. Previously this breaking was indicated for nuclei in the valley of stability in a few theoretical and experimental studies only and many other experiments were not sensitive to it. We now show its value for a description of level densities in their dependence on energy and angular momentum as obtained for nuclei in the valley of stability and this is important for predictions on compound nuclear reactions. For more than 100 spin-0 target nuclei with A>50 resonance spacing data and average capture cross sections are well described by formulae with only a surprisingly small number of freely adjusted parameters, which turn out to be global i.e. A-independent, when a breaking of axial symmetry is accepted. Its application also for exotic nuclei seems attractive and the new global ansatz derived in the present work has the potential to yield good predictions for radiative neutron capture, which are important for nuclear astrophysics and for the transmutation of nuclear waste.

### Acknowledgements

This work is supported by the German federal ministry for education and research BMBF (02NUK13A) and the Europ. Commission through Fission-2013-CHANDA (project no. 605203). Intense discussions within these projects and also with other colleagues, and especially with Ralph Massarczyk, Ronald Schwengner, Julian Srebrny and Hermann Wolter, are gratefully acknowledged.

**References**

1. A. Bohr and B. Mottelson, Nuclear Structure, ch. 6, Benjamin, Reading (Mass.) (1975)
2. S.E. Larsson, Physica Scripta, 8 (1973) 17
3. M. Girod and B.Grammaticos, Phys. Rev. C 27 (1982) 2317
4. J. Meyer-ter-Vehn, et al., Phys. Rev. Lett. 32 (1974) 383
5. K. Alder et al., Rev. Mod. Phys. 28 (1956) 432
6. A. Bohr and B.R. Mottelson, Nucl. Phys. 4 (1957) 529; id., 9 (1959) 687
7. S. Frauendorf, Rev. Mod. Phys. 73, 463 ( 2001)
8. K. Kumar, Phys. Rev. Lett. 28 (1972) 249
9. D. Cline, Ann. Rev. Nucl. Part. Sci. 36 (1986) 683
10. A. Mauthofer et al., Zeitschr. f. Phys. 336, 263 (1990)
11. C. Y. Wu and D. Cline, Phys. Rev. C 54 (1996) 2356
12. J. Srebrny et al., Nucl. Phys. A 766 (2006) 25; id., Int. J. of Mod. Phys. E 20 (2011) 422 and refs
13. V.A. Plujko et al., At. Data and Nucl. Data Tab. 97 (2011) 567; http://www-nds.iaea.org/RIPL-3
14. A.R. Junghans et al., Phys. Lett. B 670 (2008) 200
15. C. Nair et al., Phys. Rev. C 78, 055802 (2008)
16. M. Erhard et al, Phys. Rev. C 81, 034319 (2010)
17. C. Nair et al., Phys. Rev. C 81, 055806 (2010)
18. A.R. Junghans et al., Journ. Korean Phys. Soc. 59, 1872 (2010)
19. E. Grosse et al., Eur. Ph. Journ. WoC. 21, 04003 (2012); dto., 8, 02006 (2010)
20. R. Schwengner et al., Phys. Rev. C 81, 054315 (2010)
21. A.R. Junghans et al., Eur.Phys.Journ. ,WoC.146(2017) 05007,ND2016; doi: 10.1051/epjconf/201714605007
22. H. Rose, J. Nuclear Energy 5, 4 (1957)
23. M.B. Chadwick et al., Nucl. Data Sheets, 107, 12, 2931 (2006)
24. M. Salvatores and G. Palmiotti, Pr. Part. Nuc. Ph. 66, 144 (2011)
25. E. M. Burbidge et al., Rev. Mod. Phys. 29, 547 (1957)
26. F. Käppeler et al., Rep. Progr. Phys. 52, 945 (1989)
27. H. Schüler and Th. Schmidt, Zeits. f. Phys. 94 (1935) 457; id., 95 (1935) 265; id., 98 (1936) 430,
28. N.J. Stone, At. Data and Nucl. Data Tables 90 (2005) 75
29. J.P. Davidson, Rev. Mod. Phys. 37, 105 (1965
30. S. Raman et al., At. Data and Nucl. Data Tables 78 (2001) 1
31. D.L. Hill and J.A.Wheeler, Phys. Rev. 89 (1953) 1102
32. J.-P. Delaroche et al., Phys. Rev. C 81 (2010) 014303;ibid., supplemental material
33. A.S. Davydov and J.P. Fillipov, Nucl. Phys. A 8 (1958) 237
34. A. Hayashi, K. Hara and P. Ring, Phys. Rev. Lett. 53 (1984) 337
35. G. Schramm et al., Phys. Rev. C 85 (2011) 014311
36. G.F. Bertsch et al., Phys. Rev. Lett. 99, 032502 (2007)
37. M. Gell-Mann et al., Phys. Rev. 95 (1954) 1612
38. W. Kuhn, Zeits. f. Phys. 33, 408 (1925);F. Reiche and W. Thomas, Zeits. f. Phys. 34, 510 (1925)
39. W. Weise, Phys. Rev. Lett. 31 (1973) 773)
40. M.B. Chadwick et al, Phys. Rev. C 44, 814 (1991)
41. J. Ahrens, Nucl. Phys. A, 446, 229 (1985)
42. S.G. Kadmenskii, V.P. Markushev and V.I. Furman, Sov. J. Nucl. Phys. 37 (1983) 165
43. C. Fiolhais, Annals of Physics, 171 (1986) 186

44. C.B. Dover, R.H. Lemmer and F.J.W. Hahne, Ann. Phys. (N.Y.) 70, 458 (1972).
45. M. Goldhaber and E. Teller, Phys. Rev. 74 (1948) 1046
46. H. Steinwedel and H. Jensen, Phys. Rev. 79 (1950) 1019
47. W. Myers et al., Phys. Rev. C 15 (1977) 2032
48. B.L. Berman and S.C. Fultz, Rev. Mod. Phys. 47 (1975) 713,
49. B.L. Berman et al., Phys. Rev. C 34 (1986) 2201
50. M. Danos, Nucl. Phys. 5 (1958) 23
51. M. Danos and W. Greiner, Phys. Rev. 134, B 284 (1964)
52. G. Maino et al., Phys. Rev. C 30 (1984) 21014
53. S.Q. Zhang et al., Phys. Rev. C 80, 021307 (2009)
54. H.W. Barz, I. Rotter and J. Höhn, Nucl. Phys. A 275 (1977) 111; R. Wünsch, priv.comm.(2010)
55. R. Beyer et al., Int. J. of Mod. Phys. E 20 (2011) 431
56. B.A. Brown, Phys. Rev. Lett. 85, 5300 (2000)
57. E.F. Gordon and R. Pitthan, Nucl. Inst. & Meth. 145 (1977) 569
58. B. Bush and Y. Alhassid, Nucl. Phys. A 531 (1991) 27
59. G. Enders et al., Phys. Rev. Lett. 69 (1992) 249
60. S.S. Dietrich and B.L. Berman, At. Data and Nucl. Data Tables 38, 199 (1988) ; cf. [ex12]
61. http://www.nndc.bnl.gov/exfor/exfor00.htm; exfor/endf00.jsp
62. E. Grosse and A.R. Junghans, Landolt-Börnstein, New Series 25 D (2012) 4; Springer (Berlin)
63. P. Carlos et al., Nucl. Phys. A172, 437 (1971)
64. H.T. Nyhus et al., Phys. Rev. C 91 (2015) 015808
65. B.L. Berman et al., Phys. Rev. C 36 (1987) 1286
66. V.V. Varlamov et al., J. Phys. Atom. Nucl., 67 (2004) 2107; ibid. 75 (2012) 1339
67. V.V. Varlamov et al., Eur. Phys. J. A 50 (2014) 114
68. G.A. Bartholomew et al., Adv. Nucl. Phys. 7 (1973) 229
69. D. Brink, Nucl. Phys. 4, 215 (1957); id., Ph.D. thesis, Oxford (1955)
70. P. Axel, Phys. Rev. 126, 671 (1962).
71. E. Khan et al., Nucl. Phys. A694 (2001) 103
72. R. Capote et al., Nucl. Data Sheets 110 (2009) 3107; id., http://www-nds.iaea.org/RIPL-3/
73. H. Feshbach et al., Phys. Rev. 71(1947) 145
74. D.J. Hughes et al., Phys. Rev. 91(1953) 1423
75. A. Ignatyuk, RIPL-3, IAEA-TECDOC-1506 (2006); www-nds.iaea.org/RIPL-3/resonances
76. E. Grosse, A.R. Junghans and R. Massarczyk, Phys. Lett. B 739 (2014) 1
77. A. Gilbert and A.G.W. Cameron, Can. Journ. of Physics, 43(1965) 1446
78. L.D. Landau and E.M. Lifschitz, Statistical Physics, 2$^{nd}$ Ed., § 80, Moscow (1964),
79. M.K. Grossjean and H. Feldmeier, Nucl. Phys. A 444 (1985) 113
80. A. Koning et al., Nucl. Phys. A 810 (2008) 13; id., www.talys.eu
81. H. Bethe, Rev. Mod. Phys. 9 (1937) 69
82. T. Ericson, Advances in Physics, 9 (1960) 425, id., Nucl. Phys. 6 (1958) 62
83. L.G. Moretto et al., Journal of Physics 580 (2015) 012048
84. M. Guttormsen et al., Eur. Phys. J. A 51 (2015) 170
85. G. Rusev et al., Phys. Rev. Lett. 110 (2013) 022503; id., Phys. Rev. C 87 (2013) 054603
86. R. Massarczyk, et al., Phys. Rev. C 86, 014319 (2012); id, private communication (2014)
87. Tsukada et al., Nucl. Physics 78 (1966) 369
88. J.R. Huizenga et al., Nucl. Phys. A 223 (1974) 589
89. S.E. Vigdor and H.J. Karwowski, Phys. Rev. C 26 (1982) 1068
90. S. Bjørnholm, et al., Roch-conf., IAEA-STI/PUB/347 (1974) 367; ibid., SMl74/205
91. A. Ignatyuk et al., Phys. Rev. C 47 (1993) 1504
92. S.F. Mughabghab, C. Dunford, Phys. Rev. Lett. 81 (1998) 4083

93. T. von Egidy and D. Bucurescu, Phys. Rev. C 80 (2009) 054310
94. J.H.D. Jensen and J.M. Luttinger, Phys. Rev. 86 (1952) 907
94. A. Mengoni and Y. Nakajima, J. Nucl. Sci. & Tec. 31 (1994) 151
95. S.K. Kataria, V. S. Rarnamurthy, and S. S. Kapoor, Phys. Rev. C 18 (1978) 549
96. A.R. Junghans et al., Nucl. Phys. A 629 (1998) 635; ibid., A649 (1999) 214c
97. G. Hansen and A.S. Jensen, Nuclear Physics A406 (1983) 23&
97. W.D. Myers and W.J. Swiatecki, Ark. Fizik 36 (1967) 343.
98. P. Möller et. al., Phys. Rev. Lett. 95, 062501 (2006); id. At. Data Nucl. Data Tab 94 (2008) 758
99. J.M. Pearson, Hyp. Int 132, (2001), 59
100. L.G. Moretto et al., Phys. Rev. C 86 (2012) 021303
101. E. Grosse et al., Physica Scripta 24 (1981) 337
102. Y. Alhassid et al., Phys. Rev. Lett. 84, 4313 (2000)
103. M. Guttormsen et al., Phys. Rev. C88, 024307 (2013)
104. T.G.Tornyi et al., Phys. Rev. C89, 044323 (2014)
105. S. Goriely, S. Hilaire, A. Koning, Phys. Rev. C78 (2008) 064307
106. G.M. Tveten et al., Phys. Rev. C94, 025804 (2016); http://www.mn.uio.no/fysikk
107. J. Kopecky and M. Uhl, Phys. Rev. C 41, 1941 (1990) and previous work quoted there
107. A.M. Lane and J. E. Lynn, Proc. Phys. Soc. (London) A 70 (1957) 557
108. F. Bečvář et al., Phys. Rev. C 52 (1995) 1278
109. H. Utsunomiya et al., Phys. Rev. Lett. 100, 162502 (2008)
110. H. Utsunomiya et al., Phys. Rev. C 84, 055805 (2011)
111. A. C. Larsen and S. Goriely, Phys. Rev. C 82 (2010) 014318
112. U. Kneissl et al., J. Phys. G 32 (2006) R217
113. K. Heyde et al., Rev. Mod. Phys. 82, 2365 (2010)
114. F. Käppeler et al., Rev. Mod. Phys., 83 (2011) 157
115. I. Dillmann et al., Phys. Rev. C 81, 015801 (2010); id., http://www.kadonis.org
116. B. Pritychenko et al., At. Data and Nucl. Da. Tab. 96 (2010) 645; www.nndc.bnl.gov/astro